\documentclass[fleqn,12pt,twoside]{article}
\usepackage{espcrc1}

\usepackage{graphicx}

\newcommand{\AmS}{{\protect\the\textfont2
  A\kern-.1667em\lower.5ex\hbox{M}\kern-.125emS}}

\title{Negative-Parity Baryons in Quenched Anisotropic Lattice QCD}

\author{Y. Nemoto\address[RBRC]{RIKEN-BNL Research Center, Brookhaven
        National Laboratory, Upton 11973, USA}%
        \thanks{Y.N. was supported by the center-of-excellence (COE) program
        at YITP, Kyoto University in most stage of this work and
        thanks RIKEN, Brookhaven National Laboratory and the
        U.S. Department of Energy for providing the facilities essential
        for the completion of this work.},
        N. Nakajima\address[KOCHI]{Center of Medical Information Science,
        Kochi Medical School, Kochi 783-8505, Japan},
        H. Matsufuru\address[YITP]{Yukawa Institute for Theoretical Physics,
        Kyoto University, Kyoto 606-8502, Japan}
        and
        H. Suganuma\address[TIT]{Faculty of Science, Tokyo Institute of 
        Technology, Tokyo 152-8551, Japan} }
       
\begin{document}

\maketitle

\begin{abstract}
We study negative-parity baryon spectra in quenched anisotropic lattice QCD.
The negative-parity baryons are measured as the parity partner of the
ground-state baryons.
In addition to the flavor octet and decuplet baryons, we pay much attention
to the flavor-singlet negative-parity baryon as a three-quark state
and compare it with the $\Lambda(1405)$ baryon.
Numerical results of the flavor octet and decuplet negative-parity 
baryon masses are close to experimental values of lowest-lying negative-parity 
baryons, while the flavor-singlet baryon is much heavier than $\Lambda(1405)$.
This indicates that the $\Lambda(1405)$ would be a multi-quark state
such as the $N\bar{K}$ molecule rather than the flavor-singlet 3 quark state.
\end{abstract}

\section{INTRODUCTION}

In the lattice QCD Monte Carlo calculation, the spectroscopy of the ground-state hadrons in the quenched
approximation has been almost established and reproduces the experimental values within 10\% deviations.
On the other hand, the spectroscopy of the excited-state baryons is a current interesting subject in lattice QCD. 
In this study, we focus on the low-lying negative-parity baryons,
especially the parity partner of the ground-state baryons and the
flavor-singlet baryon, $\Lambda(1405)$.
$\Lambda(1405)$ is a mysterious baryon in the quark model viewpoint.
It is the lightest negative-parity baryon though it has strangeness.
Our purpose of this paper is to clarify whether a naive three
valence quark picture can explain the mass of $\Lambda(1405)$ using
the characteristics of quenched lattice QCD, 
and to investigate spectra of the parity partners of the ground-state 
baryons of the flavor octet and decuplet states\cite{Neg,Neg1}.
The other possible candidate for the $\Lambda(1405)$ state is the
$N\bar{K}$ (5-quark) state.
It is useful to employ the quenched approximation to know whether the dominant
component of $\Lambda(1405)$ is the three-quark state, because the
overlap with the three-quark and five-quark states is absent 
and hence the properties of the purely three-quark state can be singled out.

\section{LATTICE SETUP}

We employ the standard Wilson gauge action and the $O(a)$ improved
Wilson quark action with tadpole improvement.
We adopt the anisotropic lattice because high resolution in the
temporal direction makes us easy to follow the change of correlation
of heavy baryons.
We take the renormalized anisotropy as $\xi=a_\sigma/a_\tau=4$, 
where $a_\sigma$ and $a_\tau$ are the spatial and temporal lattice spacings, respectively.
The simulation is performed on three lattices for which
the parameters are well tuned and the errors are rather well
evaluated in Ref.\cite{Aniso01b}.
The sizes of the lattices are $12^3\times96\ (\beta=5.75, a_\sigma^{-1}=
1.34(6)$ GeV), $16^3\times128\ (\beta=5.95, a_\sigma^{-1}=1.099(9)$ GeV) and
$20^3\times160\ (\beta=6.10, a_\sigma^{-1}=1.871(14)$ GeV).
The scale $a_{\sigma}^{-1}$ is determined from the $K^*$ meson 
mass.
Determination of the parameters is described in Ref.\cite{Aniso01b}.

On the quark mass, we adopt four different values, 
which roughly cover around the strange quark mass.
We use the standard baryon operator listed in Table \ref{tab:op} which
survive in the nonrelativistic limit.
The negative-parity baryons can be also measured by these operators
using the parity projection.
They are nothing but the parity partners of the ground-state baryons.
In the source operator, each quark is smeared with the Gaussian function
of the width $\sim 0.4$ fm.

\begin{table}
\caption{Baryon operators. $C$ is the charge conjugate matrix.}
\label{tab:op}
\begin{tabular}{ll}
\hline
 Octet &
  $B_{\alpha}(\Sigma^0) = (C\gamma_5)_{\beta\gamma}[
      u_{\alpha}(d_{\beta}s_{\gamma} - s_{\beta}d_{\gamma}) 
    - d_{\alpha}(s_{\beta}u_{\gamma} - u_{\beta}s_{\gamma}) ]$ \\
 Octet ($\Lambda$) &
  $B_{\alpha}(\Lambda) = (C\gamma_5)_{\beta\gamma}[
      u_{\alpha}(d_{\beta}s_{\gamma} - s_{\beta}d_{\gamma})
    + d_{\alpha}(s_{\beta}u_{\gamma} - u_{\beta}s_{\gamma})
   -2 s_{\alpha}(u_{\beta}d_{\gamma} - d_{\beta}u_{\gamma}) ]$ \\
 Decuplet &
  $B_{\alpha k}(\Sigma^{*0}) = (C\gamma_k)_{\beta\gamma}[
      u_{\alpha}(d_{\beta}s_{\gamma} + s_{\beta}d_{\gamma}) 
    + d_{\alpha}(s_{\beta}u_{\gamma} + u_{\beta}s_{\gamma})
    + s_{\alpha}(u_{\beta}d_{\gamma} + d_{\beta}u_{\gamma}) ]$ \\
 Singlet  &
  $B_{\alpha}(\Lambda_1) = (C\gamma_5)_{\beta\gamma}[
      u_{\alpha}(d_{\beta}s_{\gamma} - s_{\beta}d_{\gamma}) 
    + d_{\alpha}(s_{\beta}u_{\gamma} - u_{\beta}s_{\gamma})
    + s_{\alpha}(u_{\beta}d_{\gamma} - d_{\beta}u_{\gamma}) ]$ \\
\hline
\end{tabular}
\end{table}

\section{RESULTS AND DISCUSSION}

\begin{figure}
\begin{center}
\includegraphics[width=7.9cm]{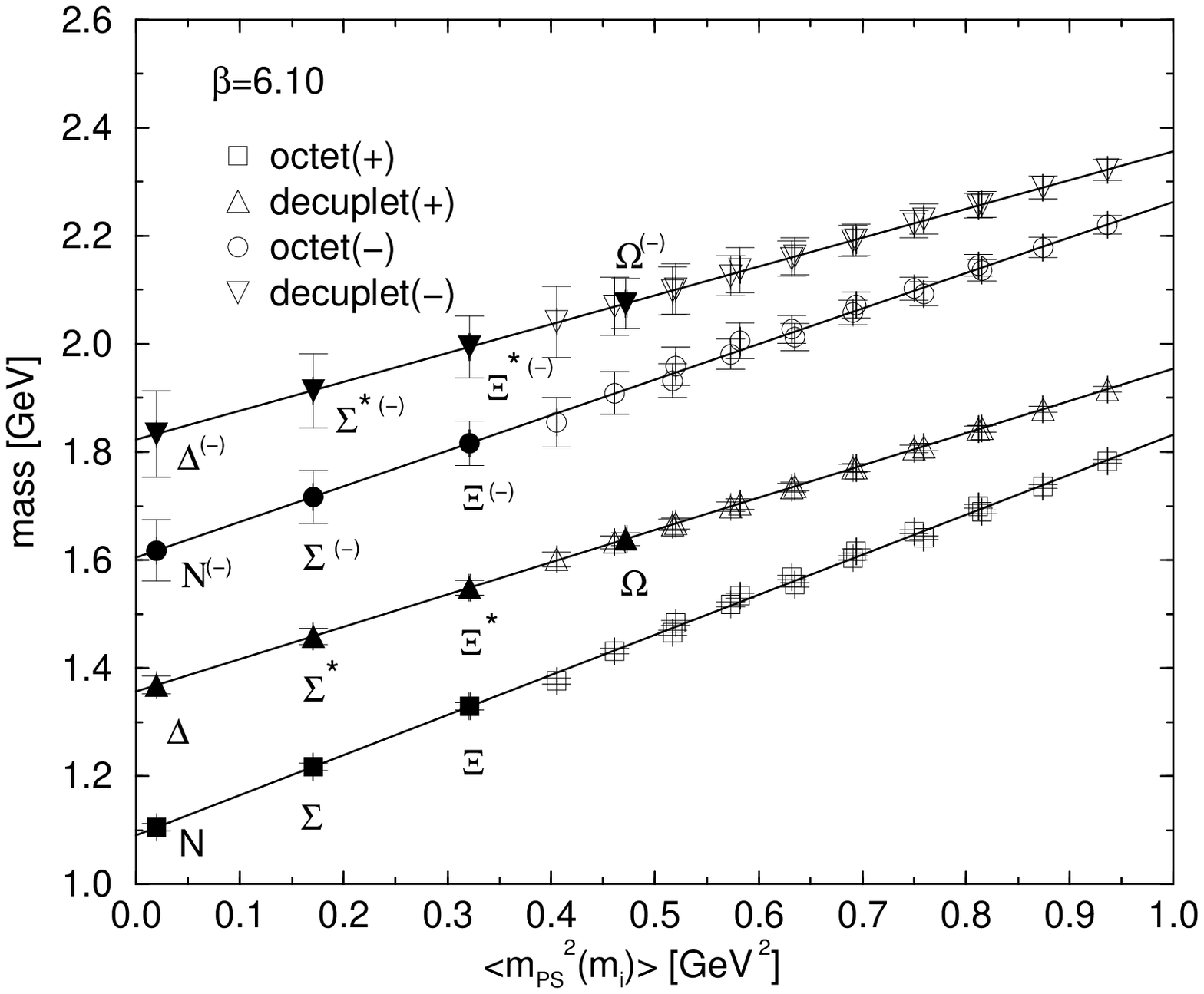}
\includegraphics[width=7.9cm]{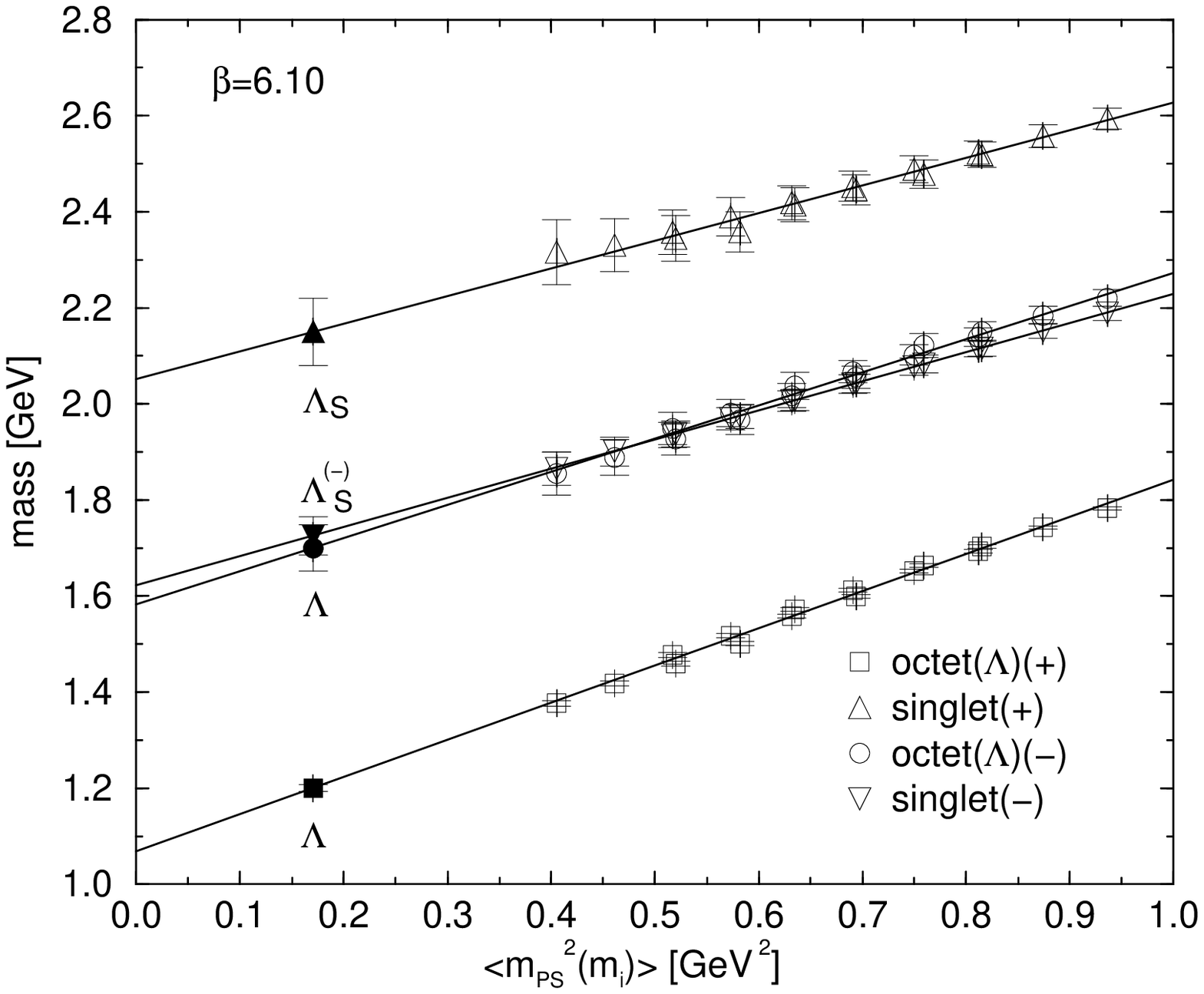}
\end{center}
\vspace{-7mm}
\caption{Spectra of the positive- and negative-parity baryons for the
$\beta=6.10$ lattice.
Octet and decuplet baryons are shown in the left figure and octet($\Lambda$)
and singlet baryons in the right figure. The open symbols denote lattice data
and the filled symbols the results fitted from the linear form of the chiral 
extrapolation.}
\label{fig:spc}
\end{figure}

The lattice data and the fit results of the baryon 
 for the $\beta=6.10$ lattice are shown in Fig. \ref{fig:spc}.
The physical $u, d$ and $s$ quark masses are determined with the $\pi$ and
$K$ meson masses.
We fit the mass data to the linear form for the pseudoscalar meson mass 
square.

\begin{figure}
\begin{center}
\includegraphics[width=6cm]{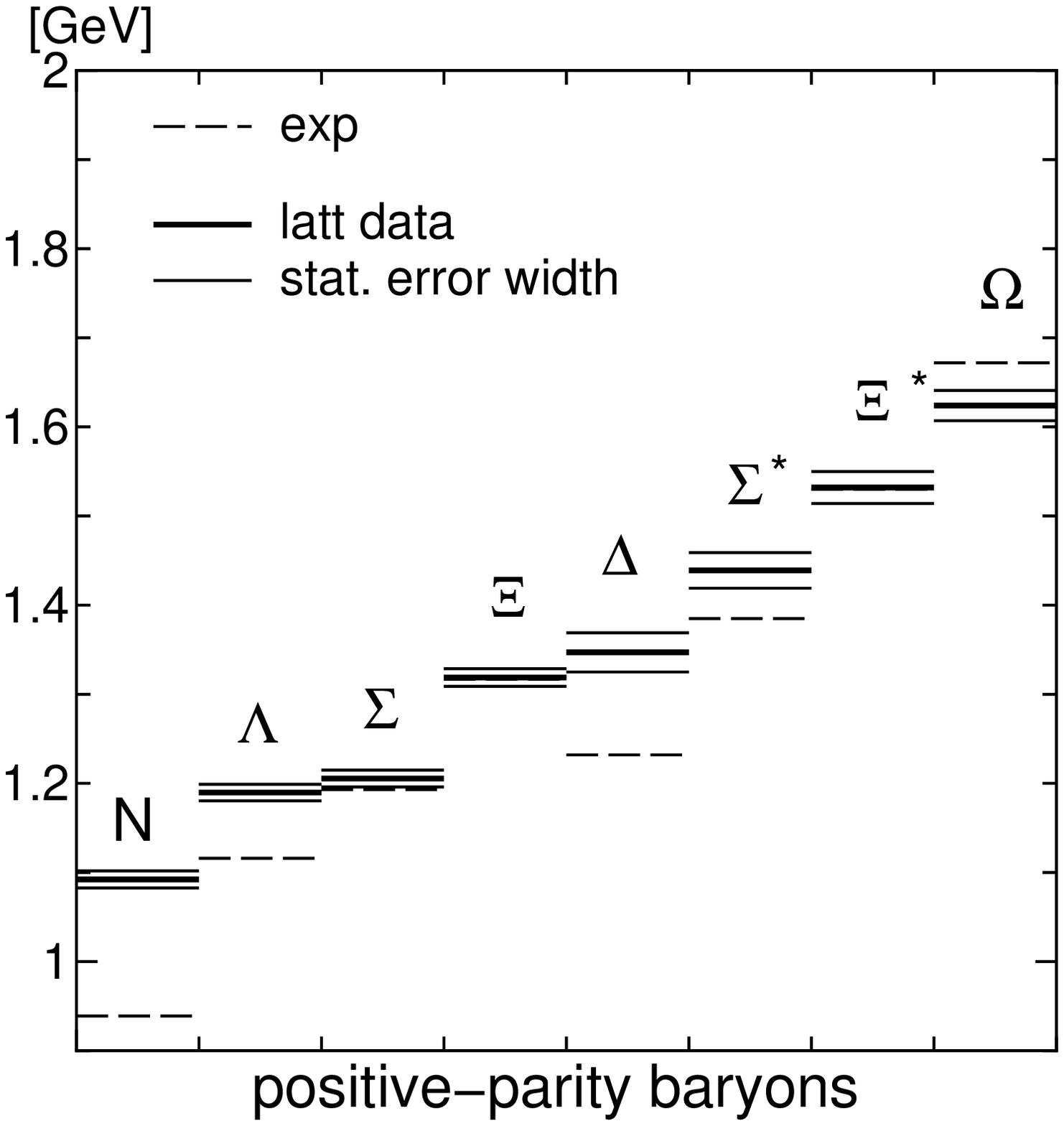}
\includegraphics[width=6.45cm]{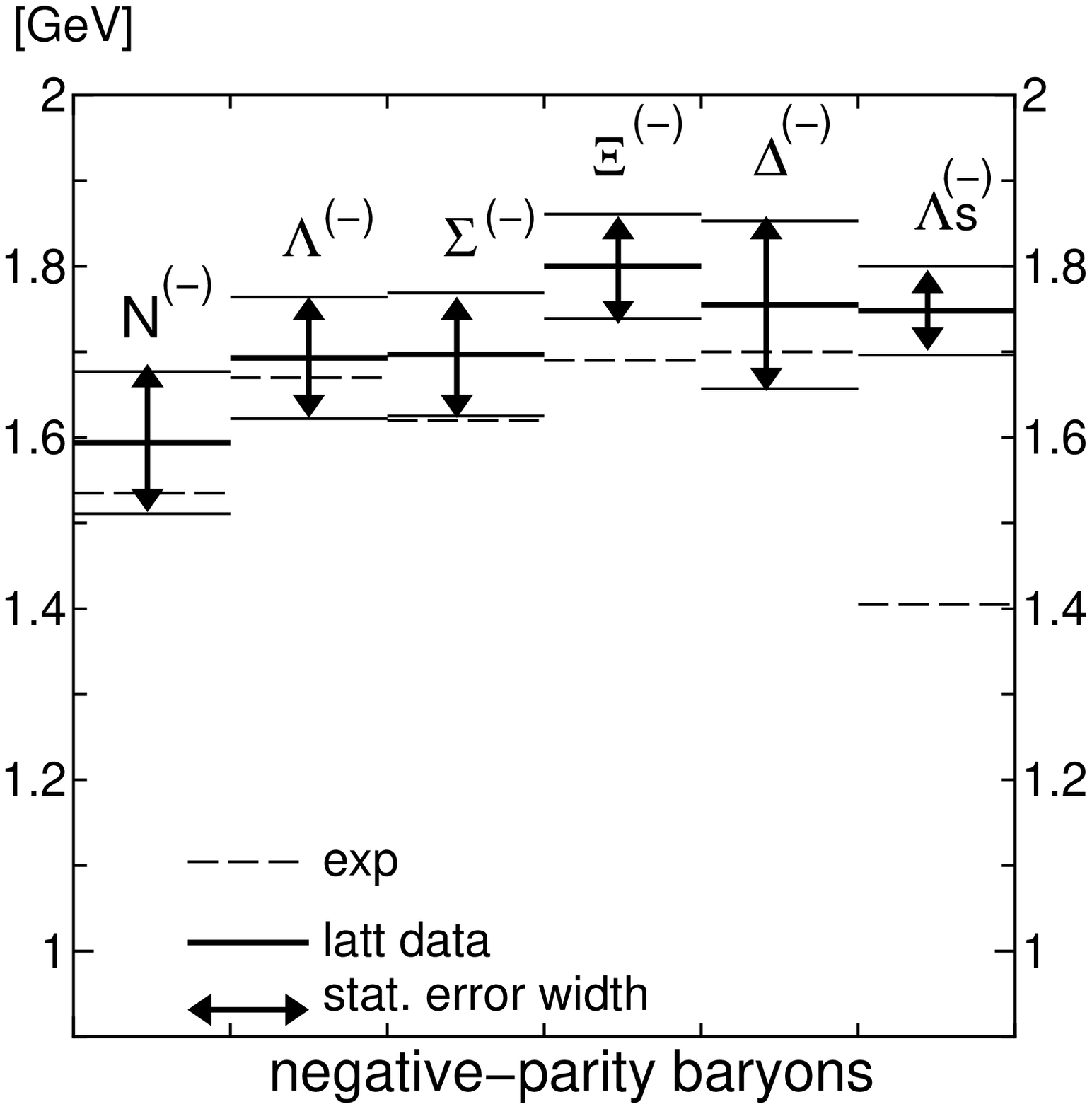}
\end{center}
\vspace{-7mm}
\caption{
Various baryon masses from the $\beta=6.10$ lattice. 
For the negative-parity baryons, we add the experimental values of   
$N(1535), \Lambda(1670), \Sigma(1620), \Xi(1690), \Delta(1700)$ and 
$\Lambda(1405)$.
}
\label{fig:spectrum4}
\end{figure}

In order to compare our lattice results with experimental values,
various baryon masses for $\beta=6.10$ lattice together with the experimental
values are shown in Fig.~\ref{fig:spectrum4}.
As for the negative-parity baryons, most of the present 
lattice results comparatively well
reproduces the experimental spectra of the lowest-lying negative-parity
baryons, in spite of  relatively large statistical errors.
(For the light baryons such as the nucleon and the delta, 
it is reported that there are non-analytic terms in the chiral
extrapolation of the nucleon in the quenched level, which 
may cause the nucleon and the delta heavier than
the experimentally observed masses to some extent.)
However, the lattice result of the flavor-singlet
negative-parity baryon is much heavier than the experimentally
observed $\Lambda(1405)$.
Note here that the flavor-singlet baryon has one strange
valence quark and therefore the ambiguity from the chiral 
extrapolation should be less than that of the nucleon and the delta.
Nevertheless, the discrepancy between the lattice result and experimental value of $\Lambda(1405)$ 
is the largest of all.
Even if we take the quenching effect of 10\% level into account,
this large discrepancy cannot be understood.
Therefore, the numerical result for the flavor-singlet baryon exhibits  
a different feature from other channels, which indicates that
it is unacceptably heavier than $\Lambda(1405)$.
Thus, we conclude that the present lattice result indicates that the three
valence quark picture
fails to represent the experimentally observed $\Lambda(1405)$, i.e.,
the overlap with the 3Q component of $\Lambda(1405)$ is small.
The other possible candidate for $\Lambda(1405)$ is a five-quark
$N\bar{K}$ state.
Lattice calculations of the $N\bar{K}$ (5Q) state will also elucidate
the nature of $\Lambda(1405)$.

We focus on other negative-parity baryons.
From Fig.~\ref{fig:spectrum4},
we see that
the lattice results of the flavor octet and decuplet sectors are all 
close to the observed lowest-lying negative parity baryons,
$N(1535), \Lambda(1670), \Sigma(1620), \Xi(1690)$ and $\Delta(1700)$,
although they have a relatively large statistical error.
The identification of the parity partner of
the sigma baryon is interesting because the evidence of existence of 
$\Sigma(1620)$ is only fair\cite{PDG02}.
As for the $\Xi$ baryons,
although the $\Xi$ with $J^P=\frac{1}{2}^-$ baryon is not known experimentally,
we can identify the parity partner as $\Xi(1690)$ from our results.
For the flavor decuplet baryons,
$\Delta(1700)$ can be regarded as the parity partner of $\Delta(1232)$.

\section{CONCLUSION}

We have studied the mass spectra of the negative-parity baryons, which 
are the parity partners of the ground-state (positive-parity) baryons,
and the flavor-singlet baryons in the quenched anisotropic lattice QCD.
The positive- and negative-parity baryon masses are extracted from the
same baryon correlators with the parity projection.

Compared with the experimental values, our present results of the
lattice calculation show that
the possible assignment of the parity partners of the ground-state
baryons for octet and decuplet is the lowest-lying negative-parity baryons.
It is also interesting to relate our lattice data to some phenomenological 
models for parity doublet baryons\cite{JNO} for future.

On the other hand,
the mass of the negative-parity flavor-singlet baryon is much heavier 
than that of
$\Lambda(1405)$, whose behavior is quite different from other negative-parity 
baryons.
This large discrepancy
between the lattice result and experimental spectrum 
seems to imply that the three valence quark picture
fails to represent the $\Lambda(1405)$ state, i.e.,
the overlap with the three-quark state of $\Lambda(1405)$ is small.
For more definite understanding of $\Lambda(1405)$ spectrum,
one needs more extensive work about the flavor-singlet
baryon such as the observation of the $N\bar{K}$ (five-quark) state.
Such a study is interesting even 
at the quenched level, where the dynamical quark-loop effect 
including the quark-antiquark pair creation 
is absent and then
the quark-level constitution of hadrons is definitely clear
in the simulation.

\end{document}